\documentclass[preprint,12pt]{elsarticle}
\usepackage{amsfonts}
\usepackage{amssymb}
\usepackage{amsmath}
\usepackage{setspace}
\usepackage{graphicx}
\usepackage{array}
\usepackage{float}
\usepackage{subfig}
\usepackage{caption}
\usepackage{hyperref}
\usepackage[utf8]{inputenc}

\makeatletter
\def\ps@pprintTitle{%
 \let\@oddhead\@empty
 \let\@evenhead\@empty
 \def\@oddfoot{}%
 \let\@evenfoot\@oddfoot}
\makeatother

\title{Study of afterpulsing in optical image intensifiers}
\author[a]{Ryan Mahon}
\author[b]{Dmitry Orlov}
\author[b]{Rene Glazenborg}
\author[a]{Andrei Nomerotski\corref{cor1}}
\cortext[cor1]{Corresponding author: anomerotski@bnl.gov}

\address[a]{Brookhaven National Laboratory, Upton NY 11973, USA}
\address[b]{Photonis Netherlands BV, 9301 ZR Roden, The Netherlands }

\hypersetup{
    colorlinks=false,
    pdfborder={0 0 0},
}

\begin{document}

\begin{abstract}
\small
We will describe the characteristics of the afterpulsing effect seen in the optical intensifiers. It can be caused by either secondary electrons produced by primary photoelectrons hitting the micro-channel plate surface or by electron emission from the photocathode induced by the ion feedback. The result of this effect are additional pulses delayed with respect to the primary parent pulses. Using a fast data-driven camera, Tpx3Cam, we were able to clearly show afterpulsing present in the data at short time differences and small distances from the primary pulse, as well as to show the evolution of the afterpulsing effect with increasing time difference. We also studied the afterpulsing spatial distribution and observed an azimuthal asymmetry, which we attribute to the afterpulsing ion component. 
\end{abstract}

\maketitle

\section{Introduction}

Optical intensifiers discussed here are vacuum devices with fast amplification, which allow for optical cameras to reach higher sensitivity. The photons are converted to photoelectrons on a photocathode and undergo avalanche amplififcation in the micro-channel plate (MCP) before they are directed to a scintillator to produce a fast flash of light. As a result, a single photon is converted to the flash, which is registered by a camera. A variety of camera types are used with intensifiers, primarily cameras based on charge-coupled devices, CCD, and on complimentary metal-oxide-semiconductor or CMOS sensors \cite{Brida2009, Reichert2017, Jost1998, Jachura2015, Fickler2013}. Here we will be focusing on single photon sensitive cameras, in particular, on the performance of intensified Tpx3Cam, a data driven camera with nanosecond scale timing resolution and fast  readout \cite{Nomerotski2019}. In our measurements we used the Tpx3Cam as a tool to investigate properties of intensifiers and, therefore, our results are applicable to all types of intensified cameras.

Below we will describe the characteristics of afterpulsing, which are the production of additional pulses delayed with respect to the primary parent pulse. Afterpulsing can mimic and distort various effects in quantum optics and quantum information science experiments, which involve registration and analysis of single photons, especially when studying the temporal and spatial correlations of multiple photons. An important example is the Hanbury Brown - Twiss (HBT) effect, which manifests itself as correlations of thermal photon pairs \cite{hbt, HanburyBrown1974, Foellmi2009}. Other examples are coincidences of single photons in the Hong-Ou-Mandel (HOM) interference \cite{Jachura2015,Nomerotski2020, Zhang2021} and false detection of bright states in ion crystals in Paul traps \cite{Zhukas2021}.

In the following sections we describe the intensified Tpx3Cam camera used for the measurements and provide details of the measurements to characterize the afterpulsing phenomenon.

\section{Tpx3Cam fast camera}
\label{sec:camera}

Imaging of single photons throughout these experiments was done with a Tpx3Cam, a data-driven time-stamping camera \cite{timepixcam, tpx3cam, Nomerotski2019, Nomerotski2023, ASI}. This camera has high quantum efficiency (QE) \cite{Nomerotski2017} and a silicon optical sensor. This sensor is then bump-bonded to Timepix3 \cite{timepix3}, which is an application specific integrated circuit (ASIC) with $256 \times 256$ pixels of $55 \times 55 $ $\mu$m$^2$. The time of arrival (ToA) is measured using the electronics present in each pixel, which processes the incoming signals for hits that cross a predefined threshold of around 600 electrons, with a precision of 1.56 ns. Another important information, measured simultaneously with ToA for each hit, is the time-over-threshold (ToT), which is related to the deposited energy in each pixel. The ToT values are stored together with the ToA values as time codes. The readout of Timepix3 is data driven, with a pixel deadtime of 475 ns + ToT, which allows for multi-hit functionality of every pixel, each independent from others and very fast at a 80 Mpix/sec bandwidth \cite{heijdenSPIDR}.  

\subsection{Intensified Camera}

For the single photon sensitive operation, the signal is amplified with the addition of Cricket$^{\rm{TM}}$ \cite{Photonis}, which contains integrated image intensifier, power supply, and relay optics to project the light flashes from the intensifier output window directly on to the optical sensor in the camera. The image intensifier is a vacuum device comprised of a photocathode followed with a micro-channel plate (MCP) and fast scintillator P47 with risetime of about 7~ns and maximum emission at 430 nm \cite{Winter2014}. The image intensifier operates at a high gain such that a single detected photon can creates a signal well above the threshold of the Timepix3 camera.  The quantum efficiency of the camera is determined primarily by the intensifier photocathode with a variety of photocathodes available, covering the spectral range from 120--900~nm\cite{Photonis, Nomerotski2019}. As an example, the hi-QE-green photocathode in the intensifier has QE of about 30\% in the range of 400 - 480~nm and hi-QE-red photocathode has QE of about 20\% in the range of 550 - 850~nm \cite{Orlov2016}. The MCP in the intensifier can have an improved detection efficiency close to 100\% \cite{Orlov2018}. 

Similar configurations of the intensified Tpx3Cam were used before for characterization of quantum networks \cite{Nomerotski2020, Ianzano2020}, quantum target detection \cite{Yingwen2020, Svihra2020}, quantum imaging \cite{Zhang2021, sensors2020, Gao2022}, studies of micromotion in ion traps \cite{Zhukas2021, Zhukas2021_1, Kato2022, Kato2023}, neutron detection \cite{DAmen2021,losko2021, Yang2021, Gao2023}, optical readout of time-projection chamber (TPC) \cite{Roberts2019, Lowe2020} and lifetime imaging \cite{Hirvonen2017, Sen2020, Sen2020_1} studies. 

\subsection{Data post-processing}

The left side of Figure \ref{fig:afterpulses} shows an example of raw TOA information for a pair of single photons registered in Tpx3Cam. Upon observation, it can be seen that photons appear as small groups of hit pixels. Subsequently, these pixels are organized into clusters within a predefined time window of 300 ns, using a recursive algorithm \cite{tpx3cam}. Since all hit pixels measure ToA and ToT independently and provide the position information, it can be used for centroiding to determine the coordinates of single photons. The ToT information is used for the weighted average, giving an estimate of the x, y coordinates for the incoming single photon with improved spatial resolution \cite{Hirvonen2017}. The timing of the photon is estimated by using ToA of the pixel with the largest ToT in the cluster.
The above ToA is then adjusted for the so-called time-walk, an effect caused by the variable pixel electronics time response, which depends on the amplitude of the input signal \cite{tpx3cam, Turecek_2016}. With this correction, a 2~ns time resolution (rms) can be achieved for single photons \cite{Ianzano2020}.

\section{Afterpulsing}
\label{sec:afterpulsing}

\begin{figure}[H]
\begin{center}
\includegraphics[width=0.46\linewidth]{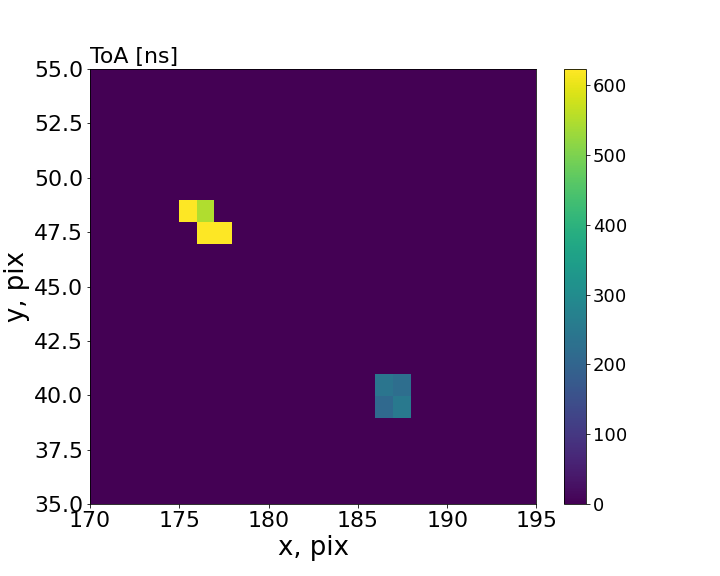}
\includegraphics[width=0.53\linewidth]{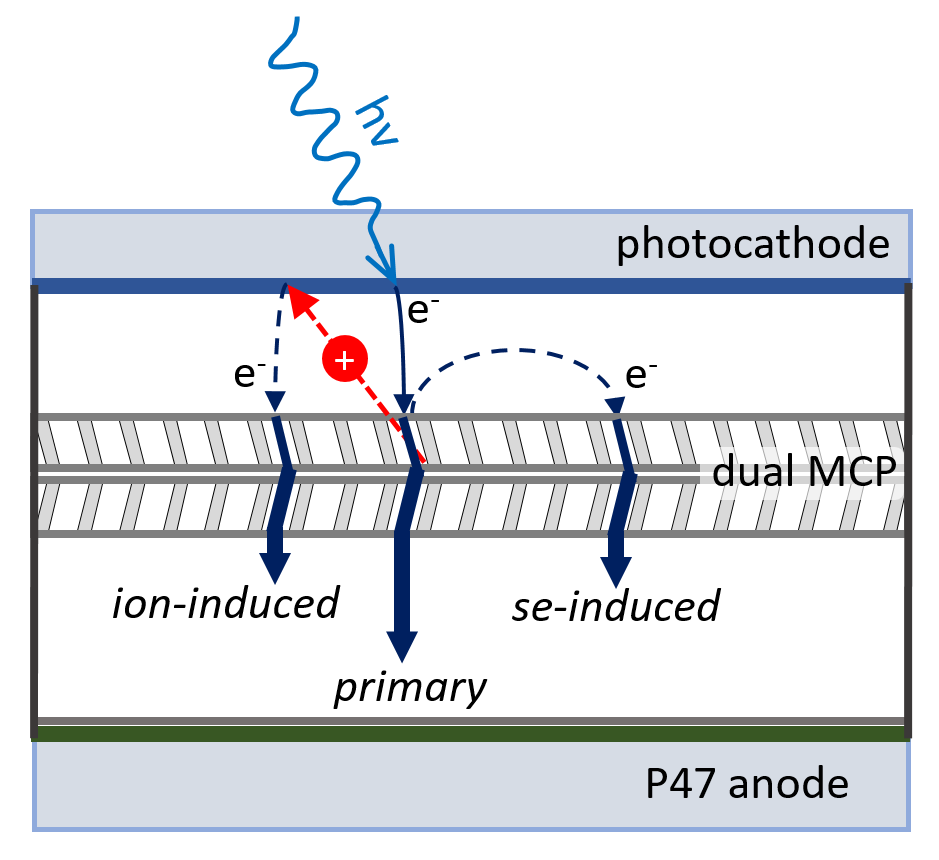}
\caption{Left: An example of raw TOA information for a pair of single photons registered in Tpx3Cam. Two clusters are seen with one delayed from the other by about 250~ns, likely caused by afterpulsing. Right: Two mechanisms of afterpulsing due to the ion feedback and secondary electrons.}
\label{fig:afterpulses}
\end{center}
\end{figure}

Alterpulsing is an important feature of the intensifier. It can be caused by secondary electrons or ions, which start another multiplication process in the vicinity of the primary pulse with a short delay  \cite{Orlov2019}. The right part of Figure \ref{fig:afterpulses} illustrates these
two mechanisms of afterpulsing generation. The first one is induced by primary photoelectrons hitting the web of the input MCP surface and producing secondary electrons. The number of secondary electrons is typically about 5 -- 10. Most of them have low energy of few eV and are quickly collected by the nearest MCP channels, thus contributing to the primary pulse. However, some of the secondary electrons have higher energy and are able to fly away from the MCP and create their own avalanches. This happens within sub-ns time and with space distribution extended to double distance between photocathode and MCP, so to about 0.2 -- 0.4~mm.   

Another source of afterpulsing is a back flow of positively charged  ions, which are generated by electron-ion stimulated desorption with the electron avalanche propagating through the MCP channels. These ions drift back to the photocathode, hitting its  surface and generating emission of photoelectrons.  Due to the ion higher mass the afterpulse is delayed with time distribution of the ion-induced afterpulses extending to 100's of ns. The left part of Figure \ref{fig:afterpulses} shows two clusters of hits resulting from two photons in a pair hitting the Tpx3Cam. There is a clear delay between these two clusters of about 300-400ns showing a clear example of the afterpulsing effect. 

Measurements described here were performed with the hi-QE-red photocathode \cite{Photonis}. We used the dark counts as a source of single photons with total rate of about 80~kHz distributed uniformly across the photocathode. This occupancy is entirely due to the thermal photoelectrons from the photocathode. After going through the registration chain (MCP-P47-Tpx3Cam), these signals look exactly the same and have same properties as response from the single photons.

\begin{figure}[H]
    \centering
    \includegraphics[width = 0.97 \linewidth]{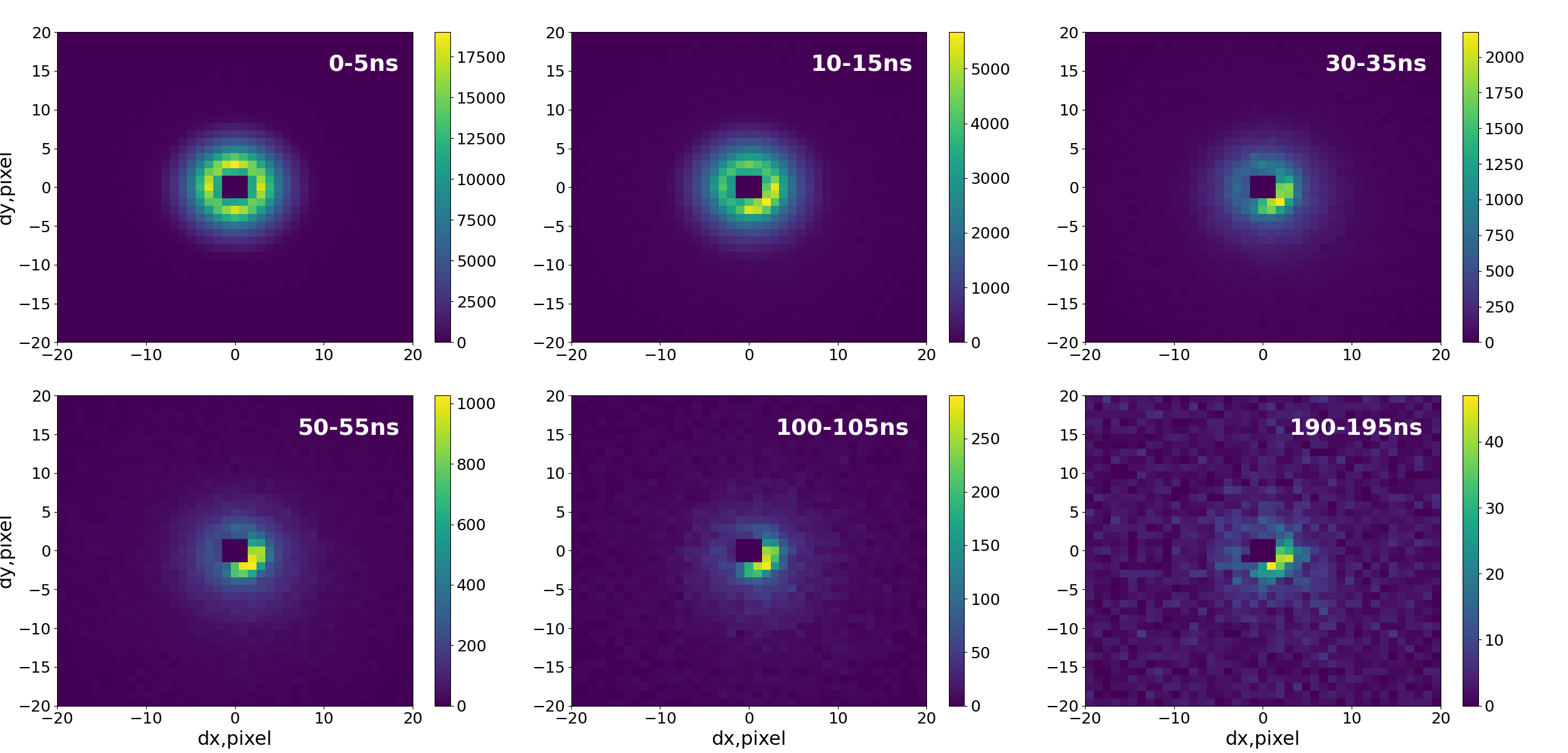}
    \caption{An evolution of the spatial distribution of afterpulses with increasing time between hits. Each graph represents a 5~ns wide slice of data with varying time difference between the pulses. The empty $3\times3$ pixel square in the middle is due to the requirement of the two hits to be separated by at least one pixel.}
    \label{fig:afterpulsing_evolution}
\end{figure}

To find the afterpulses in the data, we look at the differences in time and position between successive hits. Figure \ref{fig:afterpulsing_evolution} shows two dimensional distributions of spatial differences in both the x and y directions for increasingly large time differences between the hits. Each of six graphs in the figure represents a 5~ns wide slice of data with different time difference between the pulses. The empty $3\times3$ pixel area in the middle is due to the requirement for the two hits to be separated by at least one pixel. 
There is an obvious excess of pulses spatially close to each other due to the afterpulsing. It also can be seen from this figure that the afterpulsing effect quickly dissipates compared to its abundance in the 0-5~ns range and becomes very small (but still visible) by the time difference of about 200~ns between consecutive photon hits. 

All of these graphs, except the one corresponding to the first 5~ns bin, also have an obvious asymmetry of the azimuthal distribution, which indicates a preferred direction. This is likely related to the orientation of channels in the MCP glass with respect to the sensor coordinate system. We verified this by rotating the intensifier by 90$^\circ$ with respect to the sensor, observing that the asymmetry followed the rotation.

\begin{figure}[H]
    \centering
    \includegraphics[width = 0.42\linewidth]{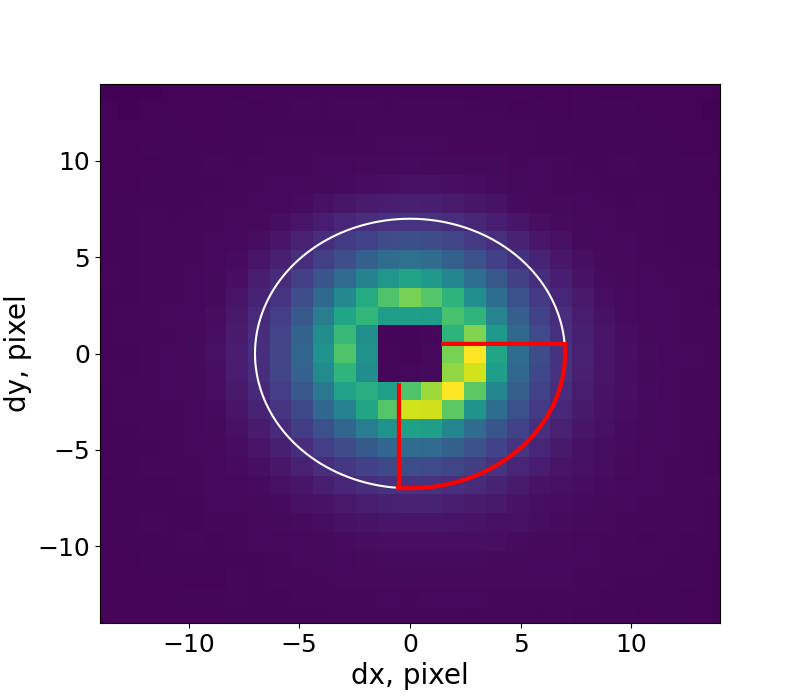}
    \includegraphics[width = 0.5\linewidth]{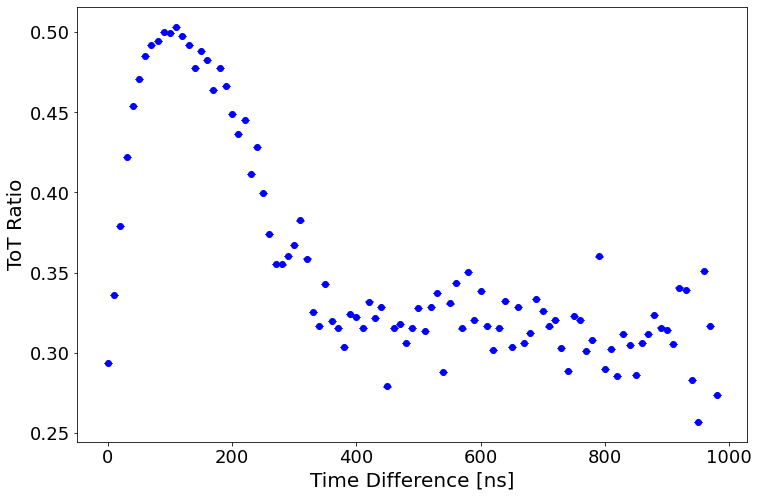}
    \caption{ Left: Diagram of spatial differences between the primary and secondary hits showing a selection of the asymmetrical region with approximate area of 30\% with respect to the total area in the circle. Right: Ratio of the sum of ToT values within asymmetrical region to the total ToT in the full area within the circle.}
    \label{fig:ToT_ratios}
\end{figure}

An interesting issue to explore was to look at the ratio of the sum of all ToT values found within the asymmetrical region compared to the full "donut" shaped region as shown in the left part of Figure \ref{fig:ToT_ratios}. The right part of the figure shows this ratio over increasing time differences from 0 to 1000~ns with binning of 10~ns. It can be seen that the ratio is about 0.3 at small time differences corresponding to the ratio of corresponding geometrical areas and indicating azimuthal uniformity.  The ratio reaches a peak of about 0.5 after 100~ns when the shape becomes asymmetric, then begins to drop off and falls back to the original 0.3 value, which is again consistent with a uniform dark count rate occupancy. This behavior suggests that the afterpulsing is only present until about 300~ns.

From the above measurements we see that there are two phases of afterpulsing: a bright and brief initial phase, which lasts  about 10~ns or less and is symmetric, and an asymmetric delayed phase with lower intensity and duration of about 300~ns. We hypothesise that the first phase is due to the secondary electrons which are produced at the MCP surface and, therefore, are symmetric and the second phase is explained by generation of ions within the MCP channels during the multiplication process. The ions are driven by electric field towards the photocathode and acquire a direction as they move along the channels. The channels have a small angle, about 7 degrees, with respect to the orthogonal direction and, we believe, it is this that creates the asymmetry in the distribution. They also have a delayed response due to their slow movement and large distance to the photocathode.

\begin{figure}[H]
\begin{center}
\includegraphics[width=0.46\linewidth]{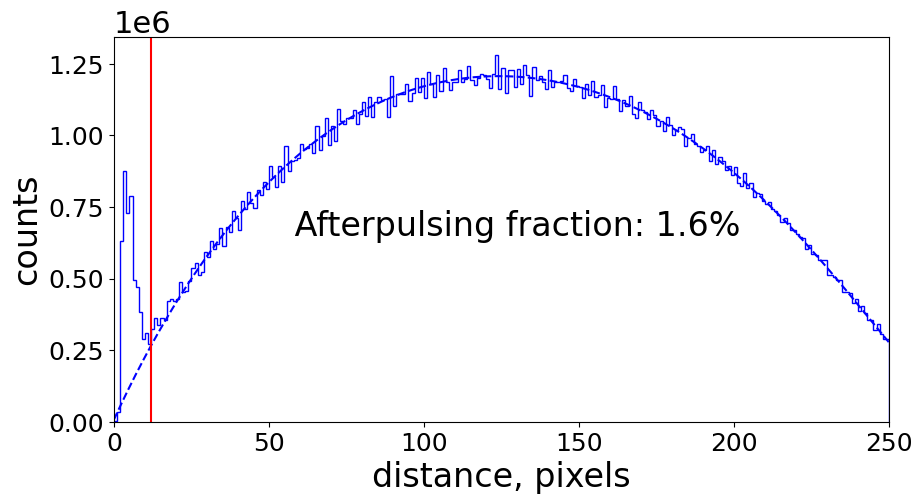}
\includegraphics[width=0.53\linewidth]{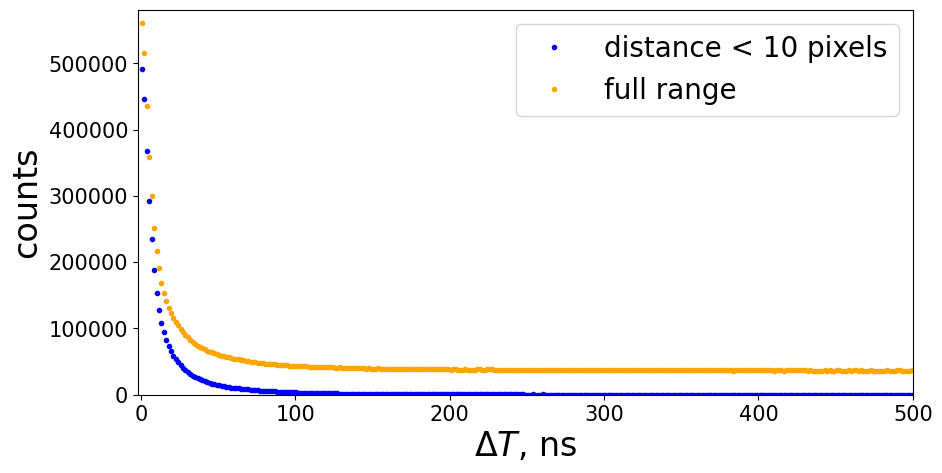}
    \caption{Left: Distribution of distances between two successive hits. Right: Distributions of time difference between successive hits for all distances and distances less than 10 pixels.}
    \label{fig:dist_fraction}
\end{center}
\end{figure}

To characterize further the afterpulsing, we plot distributions of distance between successive hits and their separation in time, see  Figure \ref{fig:dist_fraction}. There is an obvious excess for the small values in both distributions, which we estimate as deviations from the expected smooth behavior and interpret as afterpulses. We found that the fraction of afterpulsing events closer than 10 pixels is 1.6\%.  The right side of Figure \ref{fig:dist_fraction} shows distributions of time differences between successive hits for two separate cases. The first case considers all distance values. The flat background in this case corresponds to the random coincidences between the dark count rate 
(DCR) hits. The second distribution considers time differences only for cases where the corresponding distance between successive hits is less than 10 pixels. It can be seen here that the distribution drops to very small values after about 100~ns with a fraction of events closer than 100~ns to be 1.4\%. This is about the same number as obtained earlier for the spatial distribution.

\begin{figure}[H]
    \centering
    \includegraphics[width = 0.6\linewidth]{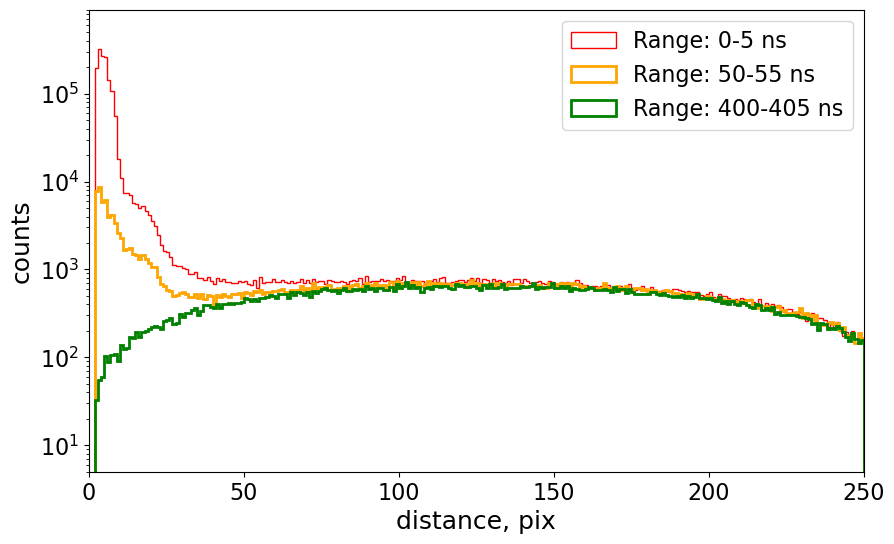}
    \caption{Distribution of distances for various time difference bins.}
    \label{fig:distance_binnings}
\end{figure}

Figure \ref{fig:distance_binnings} shows three different distributions of the distances between successive hits selected for various time difference bins. In all cases the bin width is 5~ns to maintain the same normalization at large distances, where the afterpulsing is not important. The peaks at small distance values are a result of afterpulsing and it can be seen that the peaks decrease as the time difference increases. We also note that the earliest hits in the 0 to 5~ns bin have a distribution, which is more peaked at 0 meaning smaller spatial separations between the parent hit and afterpulse in the electron phase of afterpulsing. For the last bin at 400~ns the afterpulsing is not visible confirming that the ion phase of afterpulsing ends at about 300~ns.

\begin{figure}[H]
\begin{center}
\includegraphics[width=0.48\linewidth]{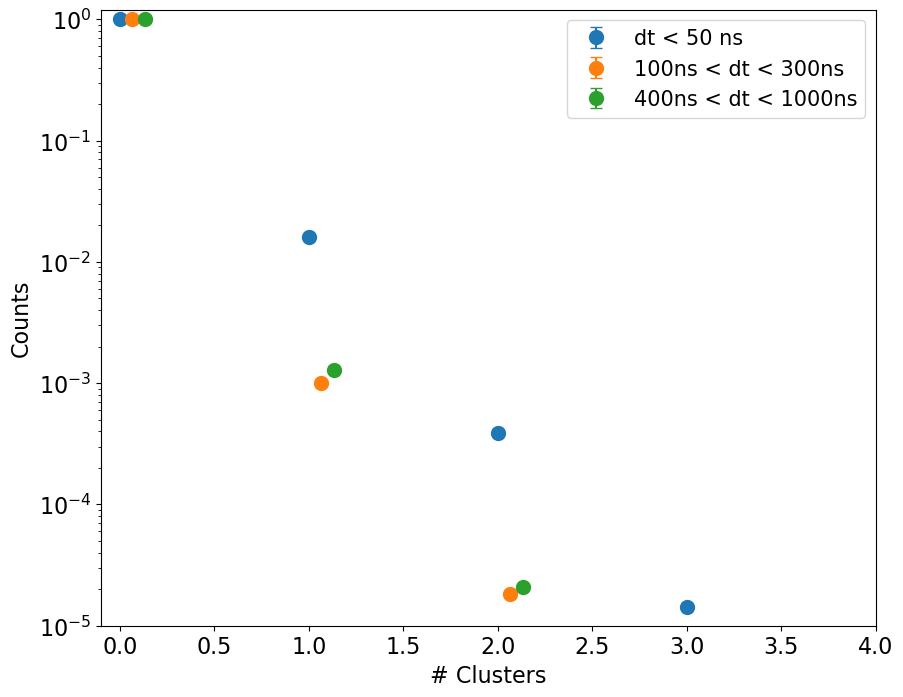}
\includegraphics[width=0.48\linewidth]{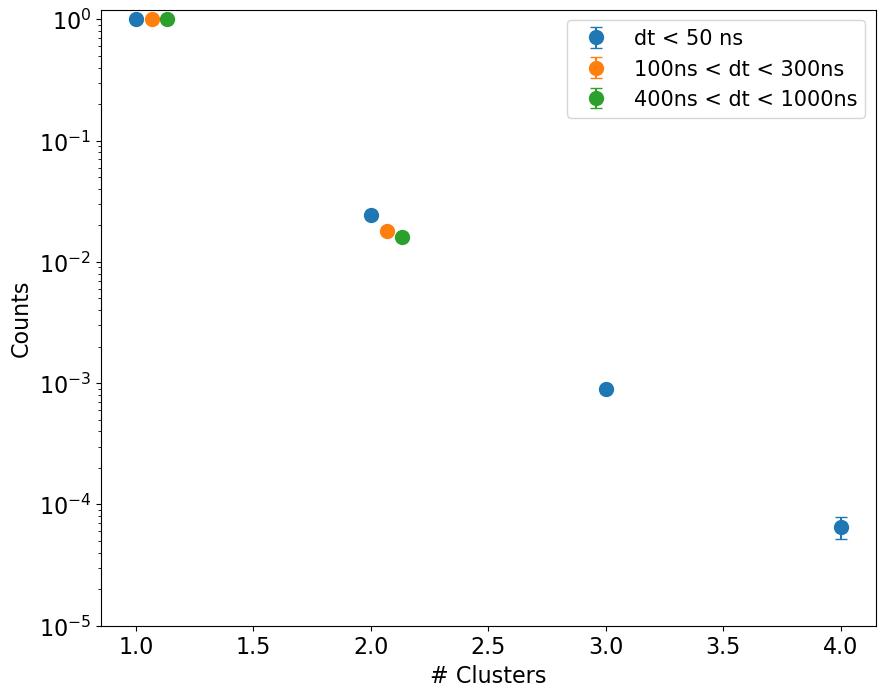}
    \caption{Left: Distribution of number of clusters,  including zero, in the 1000~ns time window after the parent pulse. Right: Distribution of number of clusters,  excluding zero, in the 1000~ns time window after the parent pulse. The distributions are normalized to the first bin.}
    \label{fig:num_clusters}
\end{center}
\end{figure}

At last, we studied the multiplicity of afterpulses. The left plot in Figure \ref{fig:num_clusters} shows a distribution of number of clusters found after the primary pulse, normalized to the number of counts at zero, which corresponds to the case of no afterpulses. These distributions represent the number of clusters found within a 1000~ns time period after an initial cluster and within a predefined distance window, from 0 to 20 pixels. Again, it illustrates the fact that afterpulses are dominated by the hits close in time and space to the original parent hit. The right plot shows the same information but it starts at one so excluding zero hits and requiring presence of at least one afterpulse. These distributions show that there is no big difference in production of multiple afterpulses between different mechanisms, that would be distinguishable in the Tpx3Cam camera.

\section{Conclusions}

We studied the afterpulsing effect seen in the optical intensifiers using a fast data-driven camera, Tpx3Cam.  We were able to clearly show afterpulsing present in the data at short time differences and distances from the primary pulse, as well as show the evolution of the afterpulsing effect with increasing time differences. We also studied the afterpulsing spatial distribution, which shows a considerable azimuthal asymmetry. We demonstrated different properties of afterpulsing for two different phases of this effect attributing them to either secondary electron or ion mechanism. Performance of the fast Tpx3Cam camera was the key element for the measurement, providing necessary temporal and spatial information about the afterpulses, which was necessary to distinguish different mechanism contributions.

The results of these studies provide useful insights for challenging measurements of photon coincidences with intensified cameras, which may also have spatial correlations. As an example we refer to the measurement of bunched photons due the HOM effect reported in \cite{Nomerotski2020} and note that the authors had to introduce a 28\% correction to account for the afterpulsing, while the probability for a single photon to have an afterpulse, which would pass the analysis selections was 0.19\%. We note that in this and all similar cases the afterpulsing effect can be evaluated by collecting independent datasets and by subtracting the afterpulsing contribution to the expected signal.

\bibliographystyle{unsrt}
\bibliography{NIM_afterpulsing}

\end{document}